\documentclass[entropy,article,submit,oneauthor,LaTeX]{Definitions/mdpi} 
\usepackage[english]{babel}
\usepackage{rotating,multirow}
\usepackage{bm}
\usepackage{multirow,mathrsfs}
\usepackage{amssymb}
\usepackage{cuted}

\newcommand{\AJP}{ {\em Am. J. Phys.} }
\newcommand{\AM}{ {\em Annals Math.} }

\newcommand{\APB}{ {\em Ann. Phys. (Berlin)} }
\newcommand{\APNY}{ {\em Ann. Phys. (N.Y.)} }

\newcommand{\CMP}{ {\em Commun. Math. Phys.} }

\newcommand{\CPL}{ {\em Chem. Phys. Lett.} }
\newcommand{\CRA}{ {\em C. R. Acad. Sci. Ser. A} }
\newcommand{\EJP}{ {\em Eur. J. Phys.} }
\newcommand{\EPJB}{ {\em Eur. Phys. J. B} }

\newcommand{\EPL}{ {\em Europhys. Lett.} }
\newcommand{\IJMPA}{ {\em Int. J. Mod. Phys. A} }
\newcommand{\IJMPB}{ {\em Int. J. Mod. Phys. B} }

\newcommand{\IJQC}{ {\em Int. J. Quantum Chem.} }
\newcommand{\JAP}{ {\em J. Appl. Phys.} }
\newcommand{\JCP}{ {\em J. Chem. Phys.} }
\newcommand{\JHEP}{ {\em J. High Energy Phys.} }
\newcommand{\JMC}{ {\em J. Math. Chem.} }
\newcommand{\JSP}{ {\em J. Stat. Phys.} }

\newcommand{\JMP}{ {\em J. Math. Phys.} }

\newcommand{\NJP}{ {\em New J. Phys.} }
\newcommand{\NL}{ {\em Nature (London)} }
\newcommand{\NLe}{ {\em Nano Lett.} }
\newcommand{\PA}{ {\em Physica A} }

\newcommand{\PLA}{ {\em Phys. Lett. A} }
\newcommand{\PNAS}{ {\em P. Natl. Acad. Sci. USA} }
\newcommand{\PR}{ {\em Phys. Rev.} }
\newcommand{\PRA}{ {\em Phys. Rev. A} }
\newcommand{\PRB}{ {\em Phys. Rev. B} }

\newcommand{\PRE}{ {\em Phys. Rev. E} }
\newcommand{\PRL}{ {\em Phys. Rev. Lett.} }

\newcommand{\PU}{ {\em Phys. - Usp.} }

\newcommand{\RMP}{ {\em Rev. Mod. Phys.} }

\newcommand{\SST}{ {\em Semicond. Sci. Technol.} }

\newcommand{\UFN}{ {\em Usp. Fiz. Nauk} }

\firstpage{1} 
\pubvolume{xx}
\issuenum{1}
\articlenumber{5}
\pubyear{2019}
\copyrightyear{2019}
\history{Received: date; Accepted: date; Published: date}

\Title{R\'{e}nyi and Tsallis entropies of the Aharonov-Bohm ring in uniform magnetic fields}

\Author{O. Olendski $^{1,\dagger}$\orcidA{0000-0001-7891-1793}}

\AuthorNames{O. Olendski}

\address{%
$^{1}$ \quad Department of Applied Physics and Astronomy, University of Sharjah, P.O. Box 27272, Sharjah, United Arab Emirates; e-mail: oolendski@sharjah.ac.ae}

\abstract{One-parameter functionals of the R\'{e}nyi $R_{\rho,\gamma}(\alpha)$ and Tsallis $T_{\rho,\gamma}(\alpha)$ types are calculated both in the position (subscript $\rho$) and momentum ($\gamma$) spaces for the azimuthally symmetric 2D nanoring that is placed into the combination of the transverse uniform magnetic field $\bf B$ and the Aharonov-Bohm (AB) flux $\phi_{AB}$ and whose  potential profile is modelled by the superposition of the quadratic and inverse quadratic dependencies on the radius $r$. Position (momentum) R\'{e}nyi entropy depends on the field $B$ as a negative (positive) logarithm of $\omega_{eff}\equiv\left(\omega_0^2+\omega_c^2/4\right)^{1/2}$, where  $\omega_0$ determines the quadratic steepness of the confining potential and $\omega_c$ is a cyclotron frequency. This makes the sum ${R_\rho}_{nm}(\alpha)+{R_\gamma}_{nm}(\frac{\alpha}{2\alpha-1})$ a field-independent quantity that increases with the principal $n$ and azimuthal $m$ quantum numbers and does satisfy corresponding uncertainty relation. In the limit $\alpha\rightarrow1$ both entropies in either space tend to their Shannon counterparts along, however, different paths. Analytic expression for the lower boundary of the semi-infinite range of the dimensionless coefficient $\alpha$ where the momentum entropies exist reveals that it depends on the ring geometry, AB intensity and quantum number $m$. It is proved that there is the only orbital for which both R\'{e}nyi and Tsallis uncertainty relations turn into the identity at $\alpha=1/2$ and which is not necessarily the lowest-energy level. At any coefficient $\alpha$, the dependence of the position R\'{e}nyi entropy on the AB flux mimics the energy variation with $\phi_{AB}$ what, under appropriate scaling, can be used for the unique determination of the associated persistent current. Similarities and differences between the two entropies and their uncertainty relations are discussed too.}


\begin{document}
\section{Introduction}\label{sec1}
In an attempt to expand quantum-information theory to the study of the QRs \cite{Fomin1}, recent analysis \cite{Olendski1} addressed an influence of the combination of the transverse uniform magnetic field $\bf B$ and the AB flux ${\bm\phi}_{AB}$ \cite{Aharonov1} on the position and momentum components of, among others, Shannon entropy \cite{Shannon1} of the one-particle orbitals of the flat 2D annulus whose rotationally symmetric potential profile $U(r)$ is modelled in the position polar coordinates $(r,\varphi_r)$ by the superposition of the quadratic and inverse quadratic dependencies on the radius $r$ \cite{Bogachek1,Tan1,Tan2,Tan3,Fukuyama1,Bulaev1,Simonin1,Margulis1,Olendski2,Xiao1,Negrete1}:
\begin{equation}\label{Potential1}
U(r)=\frac{1}{2}m^*\omega_0^2r^2+\frac{\hbar^2}{2m^*r^2}\,a-\hbar\omega_0a^{1/2},
\end{equation}
where $m^*$ is an effective mass of a charge carrier, frequency $\omega_0$ defines a steepness of the confining in-plane outer surface of the QR with its characteristic radius $r_0=[\hbar/(2m^*\omega_0)]^{1/2}$, and positive dimensionless constant $a$ describes a strength of the repulsive potential near the origin. General definition of the Shannon position $S_\rho$ and momentum $S_\gamma$ quantum-information entropies  in the $l$-dimensional spaces read:
\begin{subequations}\label{Shannon1}
\begin{align}\label{Shannon1_R}
S_\rho&=-\int_{\mathbb{R}^l}\rho({\bf r})\ln\rho({\bf r})d{\bf r}\\
\label{Shannon1_P}
S_\gamma&=-\int_{\mathbb{R}^l}\gamma({\bf k})\ln\gamma({\bf k})d{\bf k},
\end{align}
\end{subequations}
with the integration carried out over the whole available regions where the corresponding waveforms $\Psi({\bf r})$ and $\Phi({\bf k})$ that enter into the densities
\begin{subequations}\label{DensityXK1}
\begin{align}\label{DensityX1}\rho({\bf r})&=\left|\Psi({\bf r})\right|^2\\
\label{DensityK1}\gamma({\bf k})&=\left|\Phi({\bf k})\right|^2,
\end{align}
\end{subequations} 
are defined. Position $\Psi({\bf r})$ and wave vector $\Phi({\bf k})$  functions are related to each other through the Fourier transformation what for the 2D geometry of the QR is expressed as:
\begin{subequations}\label{Fourier1}
\begin{align}\label{Fourier1_1}
\Phi_{nm}(k,\varphi_k)=\frac{1}{2\pi}\int_0^{2\pi}d\varphi_r\int_0^\infty drr\Psi_{nm}(r,\varphi_r)e^{-ikr\cos(\varphi_r-\varphi_k)}\\
\label{Fourier1_2}
\Psi_{nm}(r,\varphi_r)=\frac{1}{2\pi}\int_0^{2\pi}d\varphi_k\int_0^\infty dkk\Phi_{nm}(k,\varphi_k)e^{ikr\cos(\varphi_k-\varphi_r)},
\end{align}
\end{subequations}
where $(k,\varphi_k)$ are the wave vector polar coordinates and $n=0,1,2,\ldots,$ and $m=0,\pm1,\pm2,\ldots,$ are principal and azimuthal indices, respectively. Due to the rotational symmetry of the QR, either dependence is most conveniently represented as a product of the angular and radial parts:
\begin{subequations}\label{Representation1}
\begin{align}\label{Representation1_1}
\Psi_{nm}\!\left(r,\varphi_r\right)&=\frac{1}{(2\pi)^{1/2}}e^{im\varphi_r}R_{nm}(r)\\
\label{Representation1_2}
\Phi_{nm}\!\left(k,\varphi_k\right)&=\frac{(-i)^m}{(2\pi)^{1/2}}e^{im\varphi_k}K_{nm}(k),
\end{align}
\end{subequations}
where the latter ones are:
\begin{subequations}\label{RadialWaveforms1}
\begin{align}\label{RadialWaveforms1R}
&R_{nm}(r)=\frac{1}{r_{eff}}\!\!\left[\frac{n!}{\Gamma(n\!+\!\lambda\!+\!1)}\right]^{1/2}\!\!\!\!\exp\!\left(\!-\frac{1}{4}\frac{r^2}{r_{eff}^2}\!\!\right)\!\left(\!\frac{1}{2}\frac{r^2}{r_{eff}^2}\!\!\right)^{\lambda/2}\!\!\!\!L_n^\lambda\!\left(\!\frac{1}{2}\frac{r^2}{r_{eff}^2}\!\!\right)\\
\label{RadialWaveforms1K}
&K_{nm}(k)=r_{eff}\!\!\left[\frac{n!}{\Gamma(n\!+\!\lambda\!+\!1)}\right]^{1/2}\!\!\!\!\int_0^\infty\!\!\!e^{-z/2}z^{\lambda/2}L_n^\lambda(z)J_{|m|}\!\left(\!2^{1/2}r_{eff}kz^{1/2}\!\right)dz.
\end{align}
\end{subequations}
Here, $\Gamma(x)$,  $L_n^\alpha(x)$ and $J_m(x)$ are $\Gamma$-function, generalized Laguerre polynomial and $m$-th order Bessel function of the first kind, respectively \cite{Abramowitz1,Gradshteyn1}. In addition,
\begin{subequations}\label{Quantities1}
\begin{align}\label{Quantities1_1}
r_{eff}&=\left(\frac{\hbar}{2m^*\omega_{eff}}\right)^{1/2}\\
\label{Quantities1_4}
\omega_{eff}&=\left(\omega_0^2+\frac{1}{4}\omega_c^2\right)^{1/2}
\intertext{and}
\label{Quantities1_5}
\omega_c&=\frac{eB}{m^*}
\intertext{is the cyclotron frequency,}
\label{Quantities1_2}
\lambda&=\left(m_\phi^2+a\right)^{1/2}\\
\label{Quantities1_3}
m_\phi&=m+\nu
\intertext{with $\nu$ being a dimensionless AB flux, i.e., the latter one is expressed in units of the elementary flux quantum $\phi_0=h/e$:}
\label{nu1}\nu&=\frac{\phi_{AB}}{\phi_0}.
\end{align}
\end{subequations}
It is easy to check that both function sets are orthonormalized:
\begin{align}
&\int_0^{2\pi}d\varphi_r\int_0^\infty dr r\Psi_{n'm'}^*\!\left(r,\varphi_r\right)\Psi_{nm}\!\left(r,\varphi_r\right)\nonumber\\
=&\int_0^{2\pi}d\varphi_k\int_0^\infty dkk\Phi_{n'm'}^*\!\left(k,\varphi_k\right)\Phi_{nm}\!\left(k,\varphi_k\right)=\delta_{nn'}\delta_{mm'},
\end{align}
where $\delta_{nn'}$ is a Kronecker delta.

It was shown \cite{Olendski1} that Shannon position (momentum) quantum information entropy decreases (increases) with the growing field $B$ as $\pm2\ln r_{eff}$ what physically means that in the corresponding space there is more (less) information about the particle location (intensity of motion). As a result, the sum $S_\rho+S_\gamma$, describing a total amount of the simultaneous information about the charge carrier, can not be altered by the uniform magnetic component and it does always satisfy the fundamental restriction \cite{Bialynicki1,Beckner1}:
\begin{equation}\label{EntropyInequality1}
S_\rho+S_\gamma\geq l(1+\ln\pi);
\end{equation}
in particular, for our geometry this uncertainty relation becomes tight for the lowest level, $n=m=0$, of the AB-free, $\phi_{AB}=0$, QD, $a=0$: mathematically, Eqs.~\eqref{RadialWaveforms1} at the zero values of $n$, $\phi_{AB}$ and $a$ degenerate to
\begin{subequations}\label{RadialWaveforms2}
\begin{align}\label{RadialWaveforms2R}
\left.R_{0m}(r)\right|_{a=\nu=0}&=\frac{1}{r_{eff}}\frac{1}{(|m|!)^{1/2}}\!\left(\!\frac{r}{2^{1/2}r_{eff}}\!\!\right)^{|m|}\exp\!\left(\!-\frac{1}{4}\frac{r^2}{r_{eff}^2}\!\!\right)\\
\label{RadialWaveforms2K}
\left.K_{0m}(k)\right|_{a=\nu=0}&=2r_{eff}\frac{1}{(|m|!)^{1/2}}\left(2^{1/2}r_{eff}k\right)^{|m|}\exp\!\left(-r_{eff}^2k^2\right),
\end{align}
\end{subequations}
what means that at $a=\nu=0$ the functions $\Psi_{00}({\bf r})$ and $\Phi_{00}({\bf k})$ turn into the 2D Gaussians converting  relation~\eqref{EntropyInequality1} into the equality. Next, a dependence of the position entropy $S_\rho$ on the normalized AB flux $\nu$ strongly resembles that of the energy spectrum:
\begin{equation}\label{Energy0}
E_{nm}(a,\nu;\omega_c)=\hbar\omega_{e\!f\!f}\left(2n+\lambda+1\right)+\frac{1}{2}m_\phi\hbar\omega_c-\hbar\omega_0a^{1/2}.
\end{equation} 
Accordingly, the knowledge of $S_\rho-\nu$ characteristics permits to calculate the persistent current \cite{Buttiker1}, which is a negative derivative of the energy with respect to the AB intensity:
\begin{equation}\label{Curr1}
J_{nm}\equiv-\frac{e}{h}\frac{\partial E}{\partial m}\equiv-\frac{\partial E}{\partial\phi_{AB}}=-\frac{e\omega_0}{2\pi}\left[\frac{m_\phi}{\lambda}\sqrt{1+\frac{1}{4}\left(\frac{\omega_c}{\omega_0}\right)^2}+\frac{1}{2}\frac{\omega_c}{\omega_0}\right].
\end{equation}

For many years, physicists and mathematicians have been looking for and discussing generalizations of the Shannon measures from Eqs.~\eqref{Shannon1}. Notably, two most famous and frequently used outcomes of these endeavors are one-parameter functionals of the R\'{e}nyi $R_{\rho,\gamma}(\alpha)$\footnote{To tell the R\'{e}nyi entropies from the radial part of the position waveform, Eq.~\eqref{RadialWaveforms1R}, we will always write the former ones with the subscript $\rho$ or $\gamma$ denoting a corresponding space.} \cite{Renyi1,Renyi2} 
\begin{subequations}\label{Renyi1}
\begin{align}\label{Renyi1X}
R_\rho(\alpha)&=\frac{1}{1-\alpha}\ln\!\left(\int_{\mathbb{R}^l}\rho^\alpha({\bf r})d{\bf r}\right)\\
\label{Renyi1K}
R_\gamma(\alpha)&=\frac{1}{1-\alpha}\ln\!\left(\!\int_{\mathbb{R}^l}\gamma^\alpha({\bf k})d{\bf k}\right)
\end{align}
\end{subequations} 
and Tsallis $T_{\rho,\gamma}(\alpha)$ \cite{Tsallis1} (or, more correctly, Havrda-Charv\'{a}t-Dar\'{o}czy-Tsallis \cite{Havrda1,Daroczy1})
\begin{subequations}\label{Tsallis1}
\begin{align}\label{Tsallis1X}
T_\rho(\alpha)&=\frac{1}{\alpha-1}\left(1-\int_{\mathbb{R}^l}\rho^\alpha({\bf r})d{\bf r}\right)\\
\label{Tsallis1K}
T_\gamma(\alpha)&=\frac{1}{\alpha-1}\left(1-\int_{\mathbb{R}^l}\gamma^\alpha({\bf k})d{\bf k}\right)
\end{align}
\end{subequations} 
types, where a non-negative coefficient $0\leq\alpha<\infty$ controls the reaction of the system to its deviation from the equilibrium; namely, the l'H\^{o}pital's rule deduces that at $\alpha\rightarrow1$ both R\'{e}nyi and Tsallis entropies degenerate to their Shannon counterpart, Eqs.~\eqref{Shannon1}, whereas at the vanishingly small magnitudes, this parameter allows equal contributions from the random events of any frequency of actual occurrence what in the case of the infinite or semi-infinite region of integration in Eqs.~\eqref{Renyi1} and \eqref{Tsallis1} leads to the divergence of the corresponding measure (provided it exists) at $\alpha\rightarrow0$. On the other hand, extremely strong R\'{e}nyi or Tsallis parameters pick up in the corresponding probability distributions the global maxima only with a full discard of all other happenings. Using similar arguments, it can be shown that both entropies are decreasing functions of their factor $\alpha$. Let us mention another particular case of these entropies; namely, Onicescu energies \cite{Onicescu1}, or disequilibria,
\begin{subequations}\label{Onicescu1}
\begin{align}\label{Onicescu1_R}
O_\rho&=\int_{\mathbb{R}^l}\rho^2({\bf r})d{\bf r}\\
\label{Onicescu1_P}
O_\gamma&=\int_{\mathbb{R}^l}\gamma^2({\bf k})d{\bf k},
\end{align}
\end{subequations}
which describe deviations from the uniform distributions, are expressed with the help of Eqs.~\eqref{Renyi1} and \eqref{Tsallis1} as
\begin{equation}\label{Onicescu2}
O_{\rho,\gamma}\equiv e^{-R_{\rho,\gamma}(2)}\equiv1-T_{\rho,\gamma}(2).
\end{equation}
For the QR, the position (momentum) Onicescu energy increases (decreases) with the uniform field as $r_{eff}^{-2}$ $\left(r_{eff}^2\right)$ what makes the product $O_\rho O_\gamma$, similar to the sum $S_\rho+S_\gamma$, a $B$-independent quantity \cite{Olendski1}.

Sobolev inequality of the conjugated Fourier transforms \cite{Beckner2}
\begin{equation}\label{Sobolev1}
\left(\frac{\alpha}{\pi}\right)^{l/(4\alpha)}\left[\int_{\mathbb{R}^l}\rho^\alpha({\bf r})d{\bf r}\right]^{1/(2\alpha)}\geq\left(\frac{\beta}{\pi}\right)^{l/(4\beta)}\left[\int_{\mathbb{R}^l}\gamma^\beta({\bf k})d{\bf k}\right]^{1/(2\beta)}
\end{equation}
with the non-negative coefficients $\alpha$ and $\beta$ obeying the constraint:
\begin{equation}\label{RenyiUncertainty2}
\frac{1}{\alpha}+\frac{1}{\beta}=2,
\end{equation}
supplemented by the additional requirement
\begin{equation}\label{Sobolev2}
\frac{1}{2}\leq\alpha\leq1,
\end{equation}
directly establishes the uncertainty relation between the position and momentum Tsallis entropies for each bound quantum orbital \cite{Rajagopal1}:
\begin{equation}\label{TsallisInequality1}
\left(\frac{\alpha}{\pi}\right)^{l/(4\alpha)}\!\!\left[1+(1-\alpha)T_\rho(\alpha)\right]^{1/(2\alpha)}\geq\left(\frac{\beta}{\pi}\right)^{l/(4\beta)}\!\!\left[1+(1-\beta)T_\gamma(\beta)\right]^{1/(2\beta)}.
\end{equation}
Logarithmization of Eq.~\eqref{Sobolev1} yields the following inequality for the R\'{e}nyi components \cite{Bialynicki2,Zozor1}:
\begin{equation}\label{RenyiUncertainty1}
R_\rho(\alpha)+R_\gamma(\beta)\geq-\frac{l}{2}\left(\frac{1}{1-\alpha}\ln\frac{\alpha}{\pi}+\frac{1}{1-\beta}\ln\frac{\beta}{\pi}\right),
\end{equation}
for which the restriction from Eq.~\eqref{Sobolev2} is waived. Note that near its unity, the Tsallis parameter turns the corresponding uncertainty, Eq.~\eqref{TsallisInequality1}, into
\begin{equation}\label{TsallisInequality1_1}
\frac{1+\left[-2S_\rho+l(1+\ln\pi)\right](\alpha-1)/4}{\pi^{l/4}}\geq\frac{1+\left[2S_\gamma-l(1+\ln\pi)\right](\alpha-1)/4}{\pi^{l/4}},\quad\alpha\rightarrow1,
\end{equation}
what means that, first, at $\alpha=1$ it becomes an identity with each of its sides equal to dimensionless $\pi^{-l/4}$ [what follows though directly from Eq.~\eqref{Sobolev1}] and, second, due to the Beckner-Bia{\l}ynicki-Birula-Mycielski inequality \cite{Bialynicki1,Beckner1}, Eq.~\eqref{EntropyInequality1}, the relation from Eq.~\eqref{TsallisInequality1_1} turns into the strict inequality at $\alpha<1$ only, as stated above, Eq.~\eqref{Sobolev2}. At the same time, its R\'{e}nyi counterpart, Eq.~\eqref{RenyiUncertainty1}, with the help of the l'H\^{o}pital's rule degenerates in the limit $\alpha\rightarrow1$ (and, accordingly, $\beta\rightarrow1$) into its Shannon fellow, Eq.~\eqref{EntropyInequality1}. It was conjectured \cite{Olendski3} that the inequalities from Eqs.~\eqref{Sobolev1}, \eqref{TsallisInequality1} and \eqref{RenyiUncertainty1} for the lowest-energy level turn into the identities at $\alpha=1/2$. This issue will be addressed below. Important difference between the entropies is the fact that the R\'{e}nyi and Shannon functionals are additive (or extensive) whereas the Tsallis dependence is not. More information on each of the entropies can be found in many sources; see, e.g., Refs.~\cite{Jizba1,Jizba2,Tsallis2,Tsallis3}.

Unique properties of the R\'{e}nyi and Tsallis entropies explain their wide applications in almost every branch of science and other fields of human acitivity: from seismology \cite{Geilikman1} and ecology \cite{Carranza1,Rocchini1} with geography \cite{Drius1} through medicine \cite{Rosso1,Tozzi1} and biology \cite{Costa1} to quantum physics \cite{Aptekarev1,Toranzo1,Dehesa1,Aptekarev2,Nasser1,Mukherjee1,Mukherjee2,Ou1,Zeama1,Ou2}, free field theories \cite{Klebanov1,Chen1} and astronomy \cite{Dong1} with many, many others in between and beyond. Partially relevant to our discussion, let us point out that, in the latest development, very recent experiments on Bose-Einstein condensate of interacting $^{87}$Rb atoms loaded into a 1D \cite{Islam1} or 2D \cite{Kaufman1} optical lattice and on up to twenty $^{40}$Ca$^+$ ions trapped into 1D straight string \cite{Brydges1} directly measured the R\'{e}nyi entanglement entropy with $\alpha=2$ of these many-body systems. These state-of-the-art achievements open up new perspectives in probing and describing dynamics of correlated qubits and simultaneously raise new challenges for the correct theoretical description of the R\'{e}nyi and Tsallis entropies of the miscellaneous quantum structures.

In the present research, a comprehensive description of both measures is provided for the QR with the potential profile described by Eq.~\eqref{Potential1} placed into the superposition of the uniform $\bf B$ and AB $\phi_{AB}$ magnetic fields with special emphasis being paid to the derivation of the analytic results; for example, even though the expressions for the momentum components of the entropies in general can be evaluated numerically only, it is possible to get a simple formula for the lower boundary $\alpha_{TH}$ of the semi-infinite range of the R\'{e}nyi or Tsallis coefficient at which the integrals in Eqs.~\eqref{Renyi1K} and \eqref{Tsallis1K} converge. Its inspection reveals that the QD momentum functionals exist at any non-negative $\alpha$ whereas for the QR topology the threshold is determined not only by the potential (or, more precisely, by the antidot strength $a$) but also by the orbital itself. In addition, the AB flux is the only external agent that can control this boundary since $\alpha_{TH}$ does not depend on $B$. The paths along which both entropies approach at $\alpha\rightarrow1$ their Shannon counterpart are shown to be different for the R\'{e}nyi and Tsallis measures. Limiting cases   of the extremely small and infinitely large coefficient $\alpha$ are also addressed. Next, neither R\'{e}nyi nor Tsallis uncertainty relation depends on the uniform field $\bf B$. Since the lowest-orbital position $\Psi_{00}({\bf r})$  and wave vector $\Phi_{00}({\bf k})$ functions of the AB-free QD ($\nu=a=0$) are described by the 2D Gaussians, the corresponding inequalities, Eq.~\eqref{TsallisInequality1} and \eqref{RenyiUncertainty1}, are saturated for this level at any coefficient $\alpha$; in particular, for the Tsallis case a restraint from Eq.~\eqref{Sobolev2} is waived. The $n=m=0$ state is a special one also for the QR since it is the only orbital that at $\alpha=1/2$ turns Eq.~\eqref{TsallisInequality1} and \eqref{RenyiUncertainty1} into the identities. The dependence of the measures on the AB intensity is investigated too and it is shown that since the position R\'{e}nyi entropy at any coefficient $\alpha$ qualitatively repeats the energy dependence on the flux, its knowledge can be useful in predicting the associated persistent currents.

Structure of the research presented below is as follows. Measures properties in the uniform magnetic field are discussed in Sec.~\ref{sec2} where their position and momentum components are addressed first in subsections~\ref{sec2_1} and \ref{sec2_2}, respectively, whereas the uncertainty relations are studied in subsection~\ref{sec2_3}, which is divided into parts devoted to the Tsallis and R\'{e}nyi functionals. Sec.~\ref{sec3} contains an analysis of the R\'{e}nyi entropies dependence on the AB flux and its relevance to the prediction of the magnitude of the persistent currents. Discussion is wrapped up in Sec.~\ref{sec4} by some concluding remarks.
 
\section{Entropies in uniform magnetic field $\bf B$}\label{sec2}
\subsection{Position components}\label{sec2_1}
Inserting the forms of the wave functions from Eqs.~\eqref{Representation1} and \eqref{RadialWaveforms1} into the general definition of the R\'{e}nyi, Eqs.~\eqref{Renyi1}, and Tsallis, Eqs.~\eqref{Tsallis1}, entropies yields:
\begin{subequations}\label{Renyi2}
\begin{align}
&R_{\rho_{nm}}(\alpha)=2\ln r_{eff}+\ln2\pi\nonumber\\
\label{Renyi2X}
&+\frac{1}{1-\alpha}\ln\left(\left[\frac{n!}{\Gamma(n+\lambda+1)}\right]^\alpha\frac{1}{\alpha^{\alpha\lambda+1}}\int_0^\infty e^{-z}z^{\alpha\lambda}\left[L_n^\lambda\left(\frac{z}{\alpha}\right)^2\right]^\alpha dz\right)\\
&R_{\gamma_{nm}}(\alpha)=-2\ln r_{eff}+\ln2\pi+\frac{\alpha}{\alpha-1}\ln\frac{n!}{\Gamma(n+\lambda+1)}\nonumber\\
\label{Renyi2K}
&+\frac{1}{1-\alpha}\ln\int_0^\infty\!d\xi\xi\left(\left[\int_0^\infty\!e^{-z/2}z^{\lambda/2}L_n^\lambda(z)J_{|m|}\!\left(2^{1/2}\xi z^{1/2}\right)\!dz\right]^2\right)^\alpha
\end{align}
\end{subequations}
\begin{subequations}\label{Tsallis2}
\begin{align}
&T_{\rho_{nm}}(\alpha)=\frac{1}{\alpha-1}\left(1-\frac{1}{\left(2\pi r_{eff}^2\right)^{\alpha-1}}\frac{1}{\alpha^{\alpha\lambda+1}}\left[\frac{n!}{\Gamma(n+\lambda+1)}\right]^\alpha\right.\nonumber\\
\label{Tsallis2X}
&\times\left.\int_0^\infty e^{-z}z^{\alpha\lambda}\left[L_n^\lambda\left(\frac{z}{\alpha}\right)^2\right]^\alpha dz\right)\\
&T_{\gamma_{nm}}(\alpha)=\frac{1}{\alpha-1}\left[1-\left(\frac{r_{eff}^2}{2\pi}\right)^{\alpha-1}\left[\frac{n!}{\Gamma(n+\lambda+1)}\right]^\alpha\right.\nonumber\\
\label{Tsallis2K}
&\times\left.\int_0^\infty\!d\xi\xi\left(\left[\int_0^\infty\!e^{-z/2}z^{\lambda/2}L_n^\lambda(z)J_{|m|}\!\left(2^{1/2}\xi z^{1/2}\right)\!dz\right]^2\right)^\alpha\right].
\end{align}
\end{subequations}
Similar to the Shannon case \cite{Olendski1}, the whole dependence of the R\'{e}nyi position and momentum entropies on the uniform magnetic field $B$ is embedded in the terms $\pm2\ln r_{eff}$. Concerning the Tsallis functionals, a dimensional incompatibility of the two items in Eqs.~\eqref{Tsallis2} precludes their direct application for the continuous probability distributions suggesting instead the forms presented in the corresponding uncertainty relation, Eq.~\eqref{TsallisInequality1}, but below we will continue to write them keeping in mind that it is just a formal representation only.

For the ground band, $n=0$, position components can be evaluated analytically:
\begin{align}\label{Renyi3}
R_{\rho_{0m}}(\alpha)&=2\ln r_{eff}+\ln2\pi+\frac{1}{1-\alpha}\ln\frac{\Gamma(\alpha\lambda+1)}{\alpha^{\alpha\lambda+1}\Gamma^\alpha(\lambda+1)}\\
\label{Tsallis3}
T_{\rho_{0m}}(\alpha)&=\frac{1}{\alpha-1}\left[1-\frac{1}{\left(2\pi r_{eff}^2\right)^{\alpha-1}}\frac{\Gamma(\alpha\lambda+1)}{\alpha^{\alpha\lambda+1}\Gamma^\alpha(\lambda+1)}\right].
\end{align}
Three limits of the last two dependencies are:

for the R\'{e}nyi entropy:
\begin{subequations}\label{Renyi4}
\begin{align}
R_{\rho_{0m}}(\alpha)&=2\ln r_{eff}+\ln2\pi-\ln\alpha\nonumber\\
\label{Renyi4_0}
&-\left[\lambda(\gamma+\ln\alpha)+\ln\left(\alpha\Gamma(\lambda+1)\right)\right]\alpha+\ldots,\quad\alpha\rightarrow0\\
\label{Renyi4_1}
R_{\rho_{0m}}(\alpha)&=S_{\rho_{0m}}+\frac{1}{2}\lambda\left[1-\lambda\psi^{(1)}(\lambda)\right](\alpha-1)+\ldots,\quad\alpha\rightarrow1\\
R_{\rho_{0m}}(\alpha)&=2\ln r_{eff}+\ln2\pi+\lambda(1-\ln\lambda)+\ln\Gamma(\lambda+1)\nonumber\\
\label{Renyi4_2}
&+\frac{1}{\alpha}\left[\lambda(1-\ln\lambda)+\ln\Gamma(\lambda+1)+\frac{1}{2}\ln\frac{\alpha}{2\pi\lambda}\right]+\ldots,\quad\alpha\rightarrow\infty,
\end{align}
\end{subequations}
for the Tsallis entropy:
\begin{subequations}\label{Tsallis4}
\begin{align}\label{Tsallis4_0}
T_{\rho_{0m}}(\alpha)&=\frac{2\pi r_{eff}^2}{\alpha}-1+\ldots,\quad\alpha\rightarrow0\\
\label{Tsallis4_1}
T_{\rho_{0m}}(\alpha)&=S_{\rho_{0m}}+c(r_{eff},\lambda)(\alpha-1)+\ldots,\quad\alpha\rightarrow1\\
\label{Tsallis4_2}
T_{\rho_{0m}}(\alpha)&=\frac{1}{\alpha}+\ldots,\quad\alpha\rightarrow\infty,
\end{align}
\end{subequations}
where the Shannon entropy $S_{\rho_{0m}}$ is \cite{Olendski1}:$$S_{\rho_{0m}}=2\ln r_{eff}+\ln2\pi+\ln\Gamma(\lambda+1)+\lambda\left[1-\psi(\lambda)\right].$$Here, $\psi(x)=d[\ln\Gamma(x)]/dx=\Gamma'(x)/\Gamma(x)$ and $\psi^{(n)}(x)=d^n\psi(x)/dx^n$, $n=1,2,\ldots$, are psi (or digamma) and polygamma functions, respectively \cite{Abramowitz1}, and $\gamma$ is Euler's constant. Also, $c(r_{eff},\lambda)$ is a function containing a sum of several terms with miscellaneous products of different powers of $\ln r_{eff}$, $\lambda$, $\Gamma(\lambda)$, $\psi(\lambda)$ and $\psi^{(1)}(\lambda)$. Due to its unwieldy structure, we do not present its explicit form here. There are a few relevant points worth mentioning during the discussion of these equations. First, at the coefficient $\alpha$ approaching zero both position entropies diverge, Eqs.~\eqref{Renyi4_0} and \eqref{Tsallis4_0}, since, as mentioned in the Introduction, the integration of the constant value over the (semi-)infinite interval essentially yields infinity. Invoking Taylor expansion of Eqs.~\eqref{Renyi2X} and \eqref{Tsallis2X} with respect to the small parameter $\alpha$, it is easy to show that the logarithmic- and inverse-like divergences for the R\'{e}nyi and Tsallis entropies, respectively, are characteristic at $\alpha\rightarrow0$ for the arbitrary band with $n\geq1$. Second, a comparison between Eqs.~\eqref{Renyi4_1} and \eqref{Tsallis4_1} reconfirms \cite{Olendski3} that at the R\'{e}nyi and Tsallis parameter tending to unity the corresponding entropies approach their Shannon counterpart along different paths. Next, as it follows, e.g., from Eq.~\eqref{Renyi3}, at the arbitrary coefficient $\alpha$ and $\phi_{AB}=0$ the position R\'{e}nyi entropy is an increasing function of the absolute value of the azimuthal index $m$. As our numerical results show, the same statement holds true for the radial quantum number $n$ too. Also, the leading term of Eq.~\eqref{Renyi4_2} follows straightforwardly from the expression
\begin{equation}\label{RenyiInfinite1}
R_{\rho,\gamma}(\infty)=-\ln\!\left(\!\!\!\begin{array}{c}
\rho_{max}\\\gamma_{max}
\end{array}\!\!\!\right)
\end{equation}
with the subscript in the right-hand side denoting a global maximum of the corresponding function. To find its location $r_{max}$ for the position density, one needs to solve a polynomial equation
\begin{equation}\label{Trascendental1}
(\lambda-z)L_n^\lambda(z)-2zL_{n-1}^{\lambda+1}(z)=0,\quad n=0,1,\ldots,
\end{equation}
with $z=\frac{1}{2}\frac{r^2}{r_{eff}^2}$ what for the ground band reproduces the first line of Eq.~\eqref{Renyi4_2}. For adjacent higher lying set of levels with $n=1$ one has $z_{max}=\lambda+\frac{3}{2}-\frac{1}{2}\sqrt{8\lambda+9}$ and:
\begin{equation}
R_{\rho_{1m}}(\infty)=2\ln r_{eff}+\ln2\pi+\ln\Gamma(\lambda+2)+z_{max}-\lambda\ln z_{max}-2\ln\frac{\sqrt{8\lambda+9}-3}{2}.
\end{equation}
Finally, as a prerequisite to the analysis of the following subsection, let us underline that position entropies are defined at any positive R\'{e}nyi or Tsallis parameter.

\subsection{Momentum components}\label{sec2_2}
For the singly connected geometry of the QD with $a=\nu=0$ the expressions from Eqs.~\eqref{Renyi3} and \eqref{Tsallis3} simplify to
\begin{align}\label{Renyi5}
\left.R_{\rho_{0m}}(\alpha)\right|_{a=\nu=0}&=2\ln r_{eff}+\ln2\pi+\frac{1}{1-\alpha}\ln\frac{\Gamma(|m|\alpha+1)}{(|m|!)^\alpha\alpha^{|m|\alpha+1}}\\
\label{Tsallis5}
\left.T_{\rho_{0m}}(\alpha)\right|_{a=\nu=0}&=\frac{1}{\alpha-1}\left[1-\frac{1}{\left(2\pi r_{eff}^2\right)^{\alpha-1}}\frac{\Gamma(|m|\alpha+1)}{(|m|!)^\alpha\alpha^{|m|\alpha+1}}\right].
\end{align}
At the same time, with the help of Eq.~\eqref{RadialWaveforms2K} the momentum components are expressed analytically too:
\begin{align}\label{Renyi6}
\left.R_{\gamma_{0m}}(\alpha)\right|_{a=\nu=0}&=-2\ln r_{eff}+\ln\frac{\pi}{2}+\frac{1}{1-\alpha}\ln\frac{\Gamma(|m|\alpha+1)}{(|m|!)^\alpha\alpha^{|m|\alpha+1}}\\
\label{Tsallis6}
\left.T_{\gamma_{0m}}(\alpha)\right|_{a=\nu=0}&=\frac{1}{\alpha-1}\left[1-\left(\frac{2}{\pi}r_{eff}^2\right)^{\alpha-1}\frac{\Gamma(|m|\alpha+1)}{(|m|!)^\alpha\alpha^{|m|\alpha+1}}\right].
\end{align}
Note that the dependencies of the position and momentum components of, e.g., the R\'{e}nyi entropy on the coefficient $\alpha$ are, apart from the constant factor, the same what can be tracked back to the fact that the corresponding waveforms from Eqs.~\eqref{RadialWaveforms2} present modified Gaussians. This also explains why the sum of the entropies from the corresponding uncertainty relation, Eq.~\eqref{RenyiUncertainty1}, takes the same values at the R\'{e}nyi parameters of one half and infinity, see Sec.~\ref{sec2_3Renyi}.

Eqs.~\eqref{Renyi5} -- \eqref{Tsallis6} manifest that under these special conditions of the 2D singly connected topology, the momentum entropies do exist at any non-negative coefficient $\alpha$. However, situation changes drastically at $a+|\nu|\neq0$ when the topology turns into the doubly connected one. To derive the lower limit of the semi-infinite range $\left[\alpha_{TH},+\infty\right)$ where the momentum entropies exist, one needs to consider the inner integral in Eqs.~\eqref{Renyi2K} and \eqref{Tsallis2K} that, as stated before \cite{Olendski1}, does not have an analytic representation. Nevertheless, for our purpose it suffices to recall that the Laguerre polynomial $L_n^\lambda(z)$ of degree $n=0,1,2,\ldots$ is a linear combination of all powers of its argument $z$ from zero to $n$. Accordingly, considering the integral
$$
\int_0^\infty\!\!\!e^{-z/2}z^{\lambda/2+n'}J_{|m|}\!\left(\!2^{1/2}\xi z^{1/2}\!\right)dz
$$
with $n'=0,\ldots n$, one finds \cite{Gradshteyn1,Prudnikov1} that it can be represented by the Kummer confluent hypergeometric function $_1F_1(a;b;x)$ \cite{Abramowitz1,Gradshteyn1} as
\begin{align}
&\int_0^\infty\!\!\!e^{-z/2}z^{\lambda/2+n'}J_{|m|}\!\left(\!2^{1/2}\xi z^{1/2}\!\right)dz=2^{n'+1+\lambda/2}\frac{\Gamma\!\left(n'+1+\frac{\lambda+|m|}{2}\right)}{|m|!}\nonumber\\
\label{Integral1}
\times&\xi^{|m|} {_1F_1\left(n'+1+\frac{\lambda+|m|}{2};|m|+1;-\xi^2\right)}.
\end{align}
Note that for the AB-free QD the coefficient $\lambda$ simplifies to $|m|$, and then for $n'=0$ the Kummer function in Eq.~\eqref{Integral1} degenerates to the fading exponent with $\xi\equiv r_{eff}k$ recovering in this way Eq.~\eqref{RadialWaveforms2K}, as expected. In general case, replacing Laguerre polynomial in Eqs.~\eqref{Renyi2K} and \eqref{Tsallis2K} by $z^{n'}$, calculating inner integral with the help of Eq.~\eqref{Integral1} and utilizing asymptotic properties of the confluent hypergeometric function \cite{Abramowitz1}, one finds that the outer integrals in just mentioned equations will converge \cite{Fikhtengolts1} at $\alpha>1/(2+\lambda+n')$. Consequently, the upper limit of the right-hand side of this inequality, which is achieved at the smallest power of the argument of the Laguerre polynomial, $n'=0$, will determine a global range of convergence of the momentum entropies $R_\gamma$ and $T_\gamma$, and the threshold value is:
\begin{equation}\label{Threshold1}
\alpha_{TH}=\left\{\begin{array}{cc}
0,&a=\nu=0\\
\frac{1}{2+\lambda},&a+|\nu|\neq0.
\end{array}
\right.
\end{equation}
Remarkably, this range is not influenced by the uniform field $B$ being, on the other hand, a function of the potential profile, as asserted before for the 1D structures \cite{Olendski3}. Observe that Eq.~\eqref{Threshold1} contains the parameter $a$ defining the inner steepness of $U(r)$ but not the outer confinement that is characterized by $\omega_0$. Also, $\alpha_{TH}$ strongly depends on the orbital itself or, more specifically, on its azimuthal quantum number $m$, which determines the distance from the centre of the ring. In addition, recalling the definition of the parameter $\lambda$ from Eq.~\eqref{Quantities1_2}, one can use the AB flux as a switch that triggers the existence of the momentum entropies.

Next, using Eq.~\eqref{RenyiInfinite1} and the fact that for the angle-independent waveforms, $m=0$, their global maxima are achieved at the zero momentum, $k=0$, as can be easily shown from Eq.~\eqref{RadialWaveforms1K}, one calculates the corresponding densities as \cite{Prudnikov1}
\begin{equation}\label{AuxEquation2}
\gamma_{n0}({\bf 0})=r_{eff}^2\frac{2^{\lambda+1}}{\pi}\frac{n!}{\Gamma(n\!+\!\lambda\!+\!1)}\frac{\Gamma^2\left(\left[\frac{n}{2}\right]+1+\frac{\lambda}{2}\right)}{\left(\left[\frac{n}{2}\right]!\right)^2}
\end{equation}
with $[x]$ denoting an integer part of $x$, what leads to the entropies:
\begin{equation}\label{RenyiMomInfinite1}
R_{\gamma_{n0}}(\infty)=-2\ln r_{eff}+\ln\frac{\pi}{2}-\ln\!\left(2^\lambda\frac{n!}{\Gamma(n\!+\!\lambda\!+\!1)}\frac{\Gamma^2\left(\left[\frac{n}{2}\right]+1+\frac{\lambda}{2}\right)}{\left(\left[\frac{n}{2}\right]!\right)^2}\right).
\end{equation}
Note that for the AB-free QD, $a=\nu=0$, when $\lambda$ in Eq.~\eqref{RenyiMomInfinite1} turns to zero, it is consistent at $n=0$ with the limit $\alpha\rightarrow\infty$ of Eq.~\eqref{Renyi6}, as expected.

\subsection{Uncertainty relations}\label{sec2_3}
Besides playing a fundamental role in the quantum foundations, entropic uncertainty relations find miscellaneous applications in information theory, ranging from entanglement witnessing to wave-particle duality to quantum cryptography etc. \cite{Wehner1,Coles1} Below, Tsallis and R\'{e}nyi inequalities are considered separately but their common features, such as a saturation to identity, are underlined.
\subsubsection{Tsallis entropy}\label{sec2_3Tsallis}
For the ground band, $n=0$, of the singly connected topology of the QD, $a=\nu=0$, Tsallis inequality from Eq.~\eqref{TsallisInequality1} with the help of the dependencies from Eqs.~\eqref{Tsallis5} and \eqref{Tsallis6} is converted to
\begin{equation}\label{TsallisInequality2}
\left(2^{1/2}r_{eff}\right)^{\frac{1-\alpha}{\alpha}}\frac{\Gamma^{\frac{1}{2\alpha}}\left(|m|\alpha+1\right)}{\alpha^{|m|/2}\left(\pi|m|!\right)^{1/2}}\geq\left(2^{1/2}r_{eff}\right)^{\frac{\beta-1}{\beta}}\frac{\Gamma^{\frac{1}{2\beta}}\left(|m|\beta+1\right)}{\beta^{|m|/2}\left(\pi|m|!\right)^{1/2}},
\end{equation}
where the coefficients $\alpha$ and $\beta$ are conjugated by Eq.~\eqref{RenyiUncertainty2}. Obviously, due to this, Eq.~\eqref{TsallisInequality2} is dimensionally correct, as
\begin{equation}\label{RenyiUncertainty3}
\frac{1-\alpha}{\alpha}=\frac{\beta-1}{\beta}.
\end{equation}
Note that for the lowest energy orbital of this configuration, $m=0$, Eq.~\eqref{TsallisInequality2} turns into the identity at any Tsallis parameter $\alpha$ without the restriction from Eq.~\eqref{Sobolev2} what is explained by the fact that its both position and momentum probability distributions are Gaussian functions, which play a very special role for the entropic inequalities in quantum information \cite{DePalma1}. Next, as already mentioned in the Introduction, at any other azimuthal index $m$ the relation from Eq.~\eqref{TsallisInequality2} turns into the equality at $\alpha=\beta=1$ around which its dimensionless part (without the coefficient $r_{eff}$) becomes
\begin{align}
&\frac{1+\left(-\ln2-\ln(|m|!)-|m|\left[1-\psi(|m|+1)\right]\right)(\alpha-1)}{\pi^{1/2}}\geq\nonumber\\
\label{TsallisInequality3}
&\frac{1+\left(-\ln2+\ln(|m|!)+|m|\left[1-\psi(|m|+1)\right]\right)(\alpha-1)}{\pi^{1/2}},\quad\alpha\rightarrow1,
\end{align}
and since, as it follows from the properties of the psi function \cite{Abramowitz1},
\begin{equation}\label{Relation1}
\ln(|m|!)+|m|\left[1-\psi(|m|+1)\right]>0,\quad|m|\geq1,
\end{equation}
the inequality from Eq.~\eqref{TsallisInequality3} holds to the left of $\alpha=1$ only, in accordance with the general Sobolev rule, Eq.~\eqref{Sobolev2}. At the opposite side of this interval, the Tsallis relation simplifies to
\begin{equation}\label{TsallisInequality4}
r_{eff}\left(\frac{2}{\pi|m|!}\right)^{1/2}2^{|m|/2}\Gamma\left(\frac{|m|}{2}+1\right)\geq r_{eff}\left(\frac{2}{\pi|m|!}\right)^{1/2}\left(\frac{|m|}{e}\right)^{|m|/2},\,\alpha\rightarrow\frac{1}{2},
\end{equation}
where we have retained the leading terms only in the Taylor expansion of both sides of Eq.~\eqref{TsallisInequality2} around $\alpha=1/2$. The gap between the left and right sides of this inequality widens as the index $|m|$ increases. Also, at the extremely large Tsallis parameter, $\alpha\rightarrow\infty$, the dimensionless parts exchange their places and simultaneously are divided by two as compared to Eq.~\eqref{TsallisInequality4}.

\begin{figure}[H]
\centering
\includegraphics[width=0.85\columnwidth]{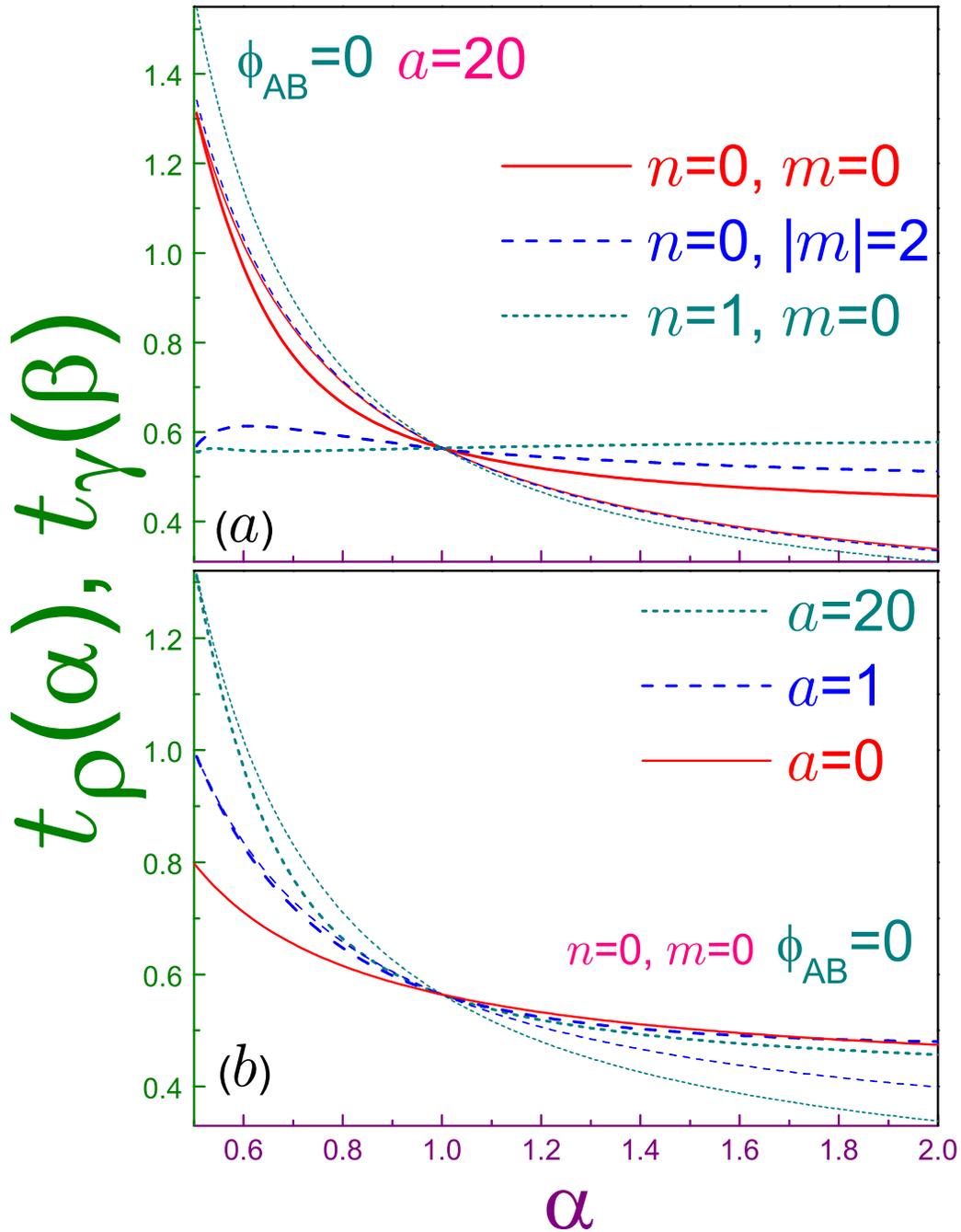}
\caption{\label{TsallisUncertaintyFig1}
Dimensionless left (thin lines) $t_{\rho_{nm}}(\alpha)$, Eq.~\eqref{tCoeffRho}, and right (thick curves) $t_{\gamma_{nm}}(\beta)$, Eq.~\eqref{tCoeffGamma}, sides of the Tsallis uncertainty relation, Eq.~\eqref{TsallisInequality1}, as functions of coefficient $\alpha$ where in panel (a) the parameter $a$ is equal to $20$ with solid lines showing the $n=m=0$ state, dashed curves are for the $n=0$, $|m|=2$ level and dotted lines stand for $n=1$, $m=0$ state whereas window (b) depicts the dependencies for several antidot strengths $a$ of the $n=m=0$ orbital: solid line is for the QD geometry ($a=0$), dashed curves are for $a=1$ and dotted functions are for $a=20$. Note that vertical ranges in parts (a) and (b) are slightly different. For both panels, the AB flux is zero.}
\end{figure}

Turning to the discussion of the general geometry of the doubly connected topology, $a+|\nu|>0$, let us note first that since here the radius $r_{eff}$ enters either side in the same way as for the QD, Eq.~\eqref{TsallisInequality2}, the Tsallis inequality at any coefficient $\alpha$ does {\em not} depend on the uniform magnetic field $\bf B$, as was the case for the Shannon regime too \cite{Olendski1}. Next, observe that at $\alpha=1/2$ the left-hand side of the general Tsallis inequality, Eq.~\eqref{TsallisInequality1}, becomes:
\begin{equation}\label{AuxEquation1}
\frac{1}{2\pi}\int_{\mathbb{R}^2}\left|\Psi_{nm}({\bf r})\right|d{\bf r}.
\end{equation}
For the rotationally symmetric orbital [when the function $\Psi_{n0}({\bf r})$ is real] of the lowest band [when the radial component $R_{0m}(r)$ preserves its sign along the $r$ axis], this expression reduces to $\Phi_{00}({\bf 0})$, see Eq.~\eqref{Fourier1_1}. On the other hand, in the same limit (i.e., at $\beta=\infty$) the right-hand sides of Eqs.~\eqref{TsallisInequality1} and \eqref{Sobolev1} turn to $$|\Phi_{nm}({\bf k})|_{max}.$$ As already mentioned in Sec.~\ref{sec2_2}, for the angle-independent, $m=0$, momentum functions their global maximum is located at the zero wave vector. Hence, we have shown that the $n=m=0$ orbital at $\alpha=1/2$ transforms the Tsallis inequality into the identity. Existence of such a level was conjectured before \cite{Olendski3} when it was stated, however, that it has to be the lowest-energy state. But the well-known property of the QR is the fact that the increasing magnetic field $\bf B$ causes consecutive crossings of the energies of the same band orbitals with adjacent non-positive azimuthal indices \cite{Tan1,Tan2,Simonin1,Olendski2}; for example, the $n=m=0$ level exhibits the lowest energy only in the range of the cyclotron frequencies from zero to \cite{Olendski2} $$2^{1/2}\frac{(a+1)^{1/2}-a^{1/2}}{\left([a(a+1)]^{1/2}-a\right)^{1/2}}\,\omega_0,$$ after which it lies above the $n=0$, $m=-1$ state. Accordingly, the previous conjecture \cite{Olendski3} stays correct in a sense that there is the only orbital that at $\alpha=1/2$ does saturate the Tsallis uncertainty relation; however, it is not necessarily the lowest-energy level (at least, for the 2D structures in the magnetic field). Solid lines in panel (a) of Fig.~\ref{TsallisUncertaintyFig1} that depicts quantities
\begin{subequations}\label{tCoeff}
\begin{align}\label{tCoeffRho}
t_\rho(\alpha)&=r_{eff}^{\frac{\alpha-1}{\alpha}}\left(\frac{\alpha}{\pi}\left[1+(1-\alpha)T_\rho(\alpha)\right]\right)^{1/(2\alpha)}\\
\label{tCoeffGamma}
t_\gamma(\beta)&=r_{eff}^{\frac{1-\beta}{\beta}}\left(\frac{\beta}{\pi}\left[1+(1-\beta)T_\gamma(\beta)\right]\right)^{1/(2\beta)},
\end{align}
\end{subequations}
which are dimensionless left and right parts, respectively, of Eq.~\eqref{TsallisInequality1}, emphasize the saturation by $n=m=0$ quantum state of the corresponding uncertainty not only at $\alpha=1$, as all other orbitals do (see dashed and dotted curves), but at the Tsallis coefficient being equal to one half too. Window (b) compares the influence of the width of the ring on the interrelation between position and momentum Tsallis parts of this orbital: it is seen that for the thinner ring (greater $a$ \cite{Olendski1,Olendski2}) the difference between them increases in the interval from Eq.~\eqref{Sobolev2}. The dependencies shown in this figure as well as in Fig.~\ref{RenyiUncertaintyFig1} are universal in a sense that they do not depend on the uniform magnetic field. For completeness, we also provide the analytic expression of the left-hand side of the Tsallis inequality for the ground band, $n=0$:
\begin{equation}\label{TsallisInequality5}
\left(2^{1/2}r_{eff}\right)^{\frac{1-\alpha}{\alpha}}\frac{\Gamma^{\frac{1}{2\alpha}}\left(\lambda\alpha+1\right)}{\pi^{1/2}\alpha^{\lambda/2}\Gamma^{\frac{1}{2}}\left(\lambda+1\right)},
\end{equation}
which generalizes its QD counterpart from Eq.~\eqref{TsallisInequality2}.
\subsubsection{R\'{e}nyi entropy}\label{sec2_3Renyi}
\begin{figure}[H]
\centering
\includegraphics[width=0.85\columnwidth]{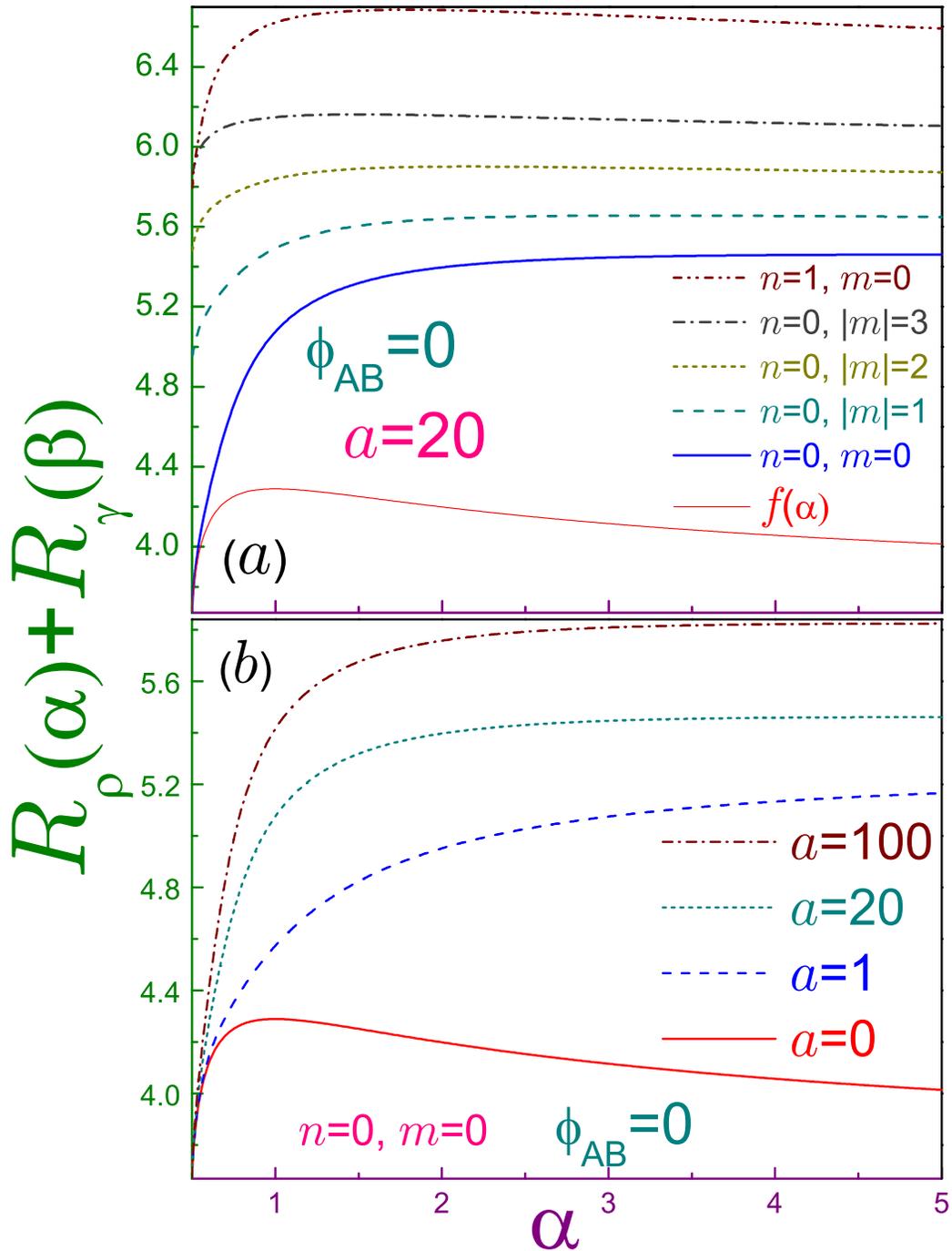}
\caption{\label{RenyiUncertaintyFig1}
Sum of the position and momentum R\'{e}nyi entropies $R_{\rho_{nm}}(\alpha)+R_{\gamma_{nm}}(\beta)$ as function of parameter $\alpha$ where in panel (a) the dependencies at the fixed antidot strength $a=20$ are shown for several indices $n$ and $m$ whereas in (b) the $n=m=0$ orbital is depicted at different $a$. In panel (a), thick solid line is for $n=m=0$ level, dotted curve - for $n=0$, $|m|=1$ state, dashed one stands for $n=0$, $|m|=2$ orbital, dash-dotted line - $n=0$, $|m|=3$ case, and dash-dot-dotted dependence describes $n=1$, $m=0$ level with thin solid curve representing function $f(\alpha)$ from Eq.~\eqref{Function_f1}, which is the right-hand side of the R\'{e}nyi uncertainty relation, Eq.~\eqref{RenyiUncertainty1}. The latter dependence is also reproduced by the solid line in panel (b) where dashed curve is for $a=1$, dotted one - for $a=20$ [corresponding to thick solid line in panel (a)], and dash-dotted curve is for $a=100$. For both windows, the AB intensity is zero, $\phi_{AB}=0$, and their upper vertical limits differ from each other.}
\end{figure}

As it follows from Eqs.~\eqref{Renyi2}, the uncertainty relation, Eq.~\eqref{RenyiUncertainty1}, is not affected by the uniform magnetic field. This statement, similar to its Tsallis counterpart from the previous subsection, expands to any R\'{e}nyi parameter a previous conclusion for the Shannon entropies \cite{Olendski1}.

Eqs.~\eqref{Renyi5} and \eqref{Renyi6} with $m=0$ directly show that the AB-free QD lowest-energy orbital saturates the entropic inequality at the arbitrary coefficient $\alpha$. Explanation of this is the same as for the Tsallis entropy, see Sec.~\ref{sec2_3Tsallis}.

For $l=2$, right-hand side of inequality~\eqref{RenyiUncertainty1}, which we will denote as
\begin{equation}\label{Function_f1}
f(\alpha)=2\left[\ln\pi-\ln\alpha+\frac{\alpha-1/2}{\alpha-1}\ln(2\alpha-1)\right],
\end{equation}
reaches its only maximum of $2(1+\ln\pi)=4.2894\ldots$ at the Shannon regime, $\alpha=1$, and approaches $2\ln2\pi=3.6757\ldots$ at $\alpha\rightarrow1/2$ and $\alpha\rightarrow\infty$ \cite{Olendski3}. For the arbitrary $m$, the same limits of the sum $R_{\rho_{0m}}(\alpha)+R_{\gamma_{0m}}(\beta)$ at $a=\nu=0$ are:
\begin{subequations}\label{HO_Flimits}
\begin{align}\label{HO_FlimitsOneHalf}
&2\ln2\pi+|m|(1+\ln2)+\ln\frac{\Gamma^2\!\left(\frac{|m|}{2}+1\right)}{|m|^{|m|}}-2\left(\alpha-\frac{1}{2}\right)\ln\left(\alpha-\frac{1}{2}\right)+\ldots,\quad\alpha\rightarrow\frac{1}{2}\\
&2(1+\ln\pi+\ln(|m|!)+|m|\left[1-\psi(|m|+1)\right])\nonumber\\
\label{HO_FlimitsOne}
&-\left[\frac{1}{3}+\frac{1}{3}|m|^3\psi(2,|m|+1)+|m|^2\psi(1,|m|+1)-\frac{2}{3}|m|\right](\alpha-1)^2+\ldots,\quad\alpha\rightarrow1\\
\label{HO_FlimitsInfinity}
&2\ln2\pi+|m|(1+\ln2)+\ln\frac{\Gamma^2\!\left(\frac{|m|}{2}+1\right)}{|m|^{|m|}}+\frac{1}{2}\frac{\ln\alpha}{\alpha}+\ldots,\quad\alpha\rightarrow\infty.
\end{align}
\end{subequations}
Note that, as it follows from Eqs.~\eqref{HO_FlimitsOneHalf} and \eqref{HO_FlimitsInfinity}, the sum of the entropies of the generalized Gaussian approaches its edge values (which are  equal to each other due to the fact that each item in it has the same dependence on the R\'{e}nyi parameter and, as a result, due to the condition from Eq.~\eqref{RenyiUncertainty2}, at the rims $\alpha$ and $\beta$ simply interchange their places) from above and since the expression in the square brackets in Eq.~\eqref{HO_FlimitsOne} is always positive, left-hand side of Eq.~\eqref{RenyiUncertainty1} reaches its maximum at the Shannon entropy. Also, the leading terms in all three cases are increasing functions of the magnetic index what means that at the greater $|m|$ the corresponding curve lies higher satisfying, of course, the uncertainty relation. As our numerical results show, the same statement holds true at the fixed quantum number $m$ and the increasing principal index $n$.

For the QR, $a>0$, a comparison of Eqs.~\eqref{Renyi3}, \eqref{RenyiMomInfinite1} and \eqref{Function_f1} proves that the $n=m=0$ orbital does convert at $\alpha=1/2$ the R\'{e}nyi uncertainty into the identity, as it did for the Tsallis inequality too. This is also exemplified in Fig.~\ref{RenyiUncertaintyFig1}(a), which shows that its sum $R_\rho(\alpha)+R_\gamma(\beta)$ at any parameter $\alpha$ is the smallest one as compared to other levels. Dependence of the left-hand side of Eq.~\eqref{RenyiUncertainty1} on $n$ and $|m|$ is the same as for the QD described in the previous paragraph. Contrary to Eqs.~\eqref{HO_FlimitsOneHalf} and \eqref{HO_FlimitsInfinity}, for the doubly connected topology the sum approaches different limits at the R\'{e}nyi parameters one half and infinity. Location of the only (relatively broad, as compared to the QD) maximum of $R_{\rho_{nm}}(\alpha)+R_{\gamma_{nm}}(\beta)$ is now $n$ and $|m|$ dependent: as panel (a) demonstrates, it is shifted to smaller $\alpha$ at the greater $|m|$ and $n$. The same effect is achieved by thinning the ring, as depicted in window (b) of Fig.~\ref{RenyiUncertaintyFig1} where also it is shown that the sum gets larger for the increasing antidot strength. In addition, it is seen that the transformation of the uncertainty relation for the $n=m=0$ state into the identity at $\alpha=1/2$ is independent from the nonzero $a$, as it follows from Eqs.~\eqref{Renyi3} and \eqref{RenyiMomInfinite1}. Finally, the remark about the conjecture \cite{Olendski3} discussed for the Tsallis entropies in Sec.~\ref{sec2_3Tsallis}, directly applies to their R\'{e}nyi counterparts too.

\section{AB R\'{e}nyi entropy}\label{sec3}
\begin{figure}[H]
\centering
\includegraphics[width=\columnwidth]{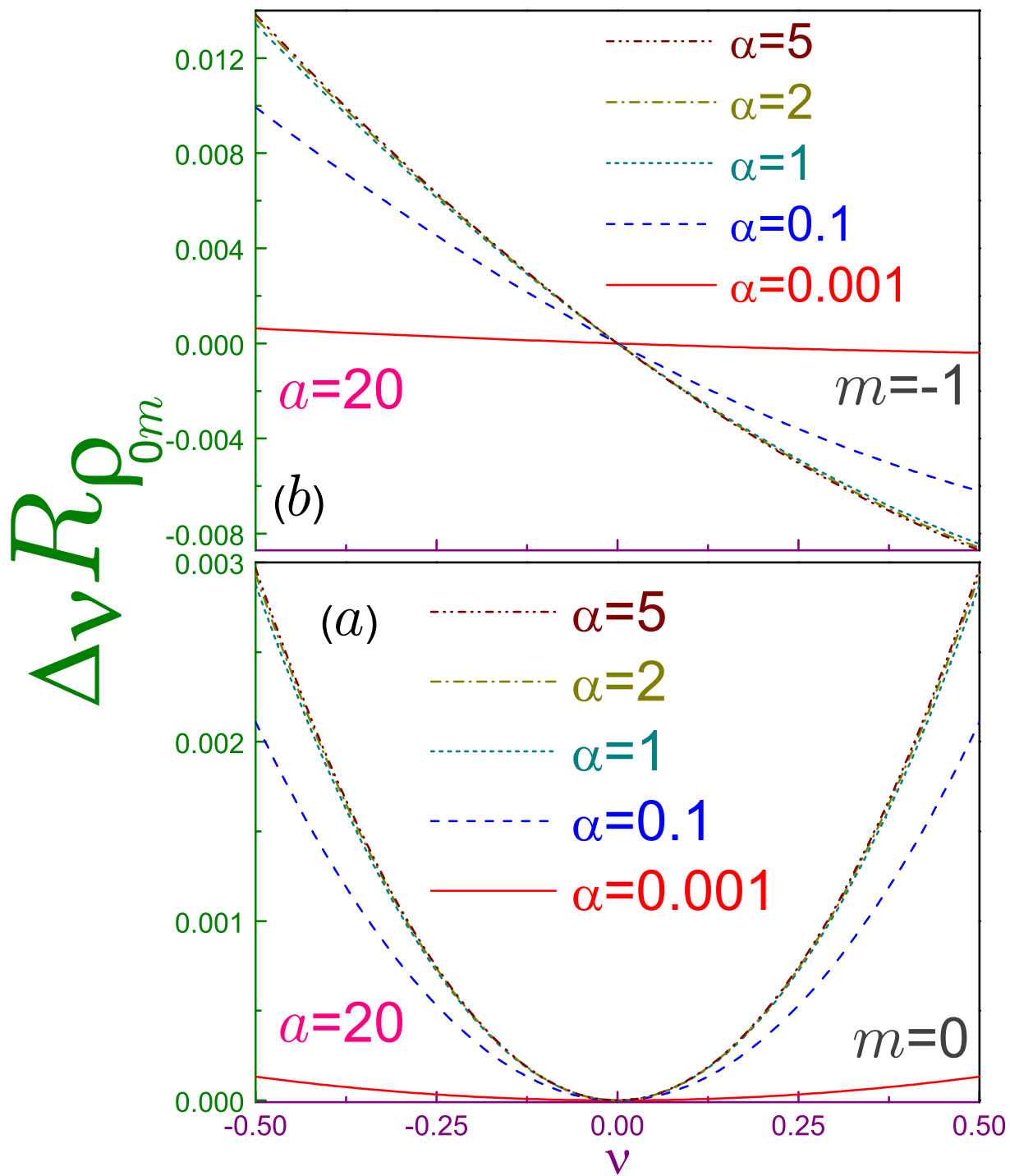}
\caption{\label{RenyiABpositionFig2}
Difference $\Delta_\nu R_{\rho_{0m}}$, Eq.~\eqref{RenyiABdiff1}, at $a=20$, $B=0$, $r_0=1$ and (a) $m=0$ and (b) $m=-1$ where solid lines are for the R\'{e}nyi parameter being equal to 0.001, dashed curves are for $\alpha=0.1$, dotted ones -- for $\alpha=1$, dashed-dotted lines -- for $\alpha=2$, and dash-dot-dotted curves depict the dependence at $\alpha=5$.}
\end{figure}

Due to the dimensional incompatibility for the continuous distributions of the two items in the right-hand sides of Eqs.~\eqref{Tsallis1}, we do not discuss  dependencies of the Tsallis measures on $\phi_{AB}$. To describe a variation of the position R\'{e}nyi entropy $R_{\rho_{00}}(\alpha)$ with the AB flux, one has to calculate Taylor expansion of Eq.~\eqref{Renyi3} with respect to the parameter $\nu$ and to truncate the series at the first nonvanishing power of the AB intensity:
\begin{align}
R_{\rho_{00}}(\nu;\alpha)=&2\ln r_{eff}+\ln2\pi+\frac{1}{\alpha-1}\left[\left(a^{1/2}\alpha+1\right)\ln\alpha+\ln\frac{\Gamma^\alpha\!\left(a^{1/2}+1\right)}{\Gamma\!\left(a^{1/2}\alpha+1\right)} \right]\nonumber\\
\label{RenyiAB1}
+&\frac{\alpha}{2a^{1/2}(\alpha-1)}\left[\psi\left(a^{1/2}+1\right)-\psi\left(a^{1/2}\alpha+1\right)+\ln\alpha\right]\nu^2.
\end{align}
Properties of digamma function \cite{Abramowitz1} applied to the analysis of the term at $\nu^2$ reveal that the entropy $R_{\rho_{00}}$ at the arbitrary R\'{e}nyi parameter and the width of the ring is, similar to the zero-uniform-field energy
\begin{equation}\label{Energy1}
E_{n0}(a,\nu;0)=\hbar\omega_0\!\left(\!2n+1+\frac{1}{2a^{1/2}}\nu^2\right),
\end{equation}
a convex function of the flux and since the persistent current is expressed with the help of the derivative of the energy with respect to $\nu$, Eq.~\eqref{Curr1}, the position entropy can be used for evaluating $J_{nm}$ too. A steepness $\partial R_{\rho_{nm}}(\nu;\alpha)/\partial\nu$ of the $R_\rho-\nu$ characteristics is strongly $\alpha-$ and $a-$dependent as, for example, three important limits show:
\begin{subequations}\label{RenyiAB2}
\begin{align}
&R_{\rho_{00}}(\nu;\alpha)=2\ln r_{eff}+\ln2\pi-\ln\alpha-\left[a^{1/2}(\gamma+\ln\alpha)+\ln\left(\alpha\Gamma\left(a^{1/2}+1\right)\right)\right]\alpha\nonumber\\
\label{RenyiAB2_0}
&-\frac{1}{2a^{1/2}}\left[\gamma+\psi\left(a^{1/2}+1\right)+\ln\alpha\right]\alpha\nu^2,\quad\alpha\rightarrow0\\
&R_{\rho_{00}}(\nu;\alpha)=2\ln r_{eff}+\ln2\pi+\ln\!\Gamma\!\left(a^{1/2}+1\right)+a^{1/2}\left[1-\psi\!\left(a^{1/2}\right)\right]\nonumber\\
&+\frac{1}{2}a^{1/2}\left[1-a^{1/2}\psi^{(1)}\left(a^{1/2}\right)\right](\alpha-1)+\frac{1}{2}\left[\frac{1}{a^{1/2}}-\psi^{(1)}\left(a^{1/2}+1\right)\right]\nu^2\nonumber\\
\label{RenyiAB2_1}
&+\left[\frac{1}{2}+\frac{3}{a^{1/2}}+a\psi^{(2)}\left(a^{1/2}\right)-a^{1/2}\psi^{(1)}\left(a^{1/2}\right)\right]\frac{\alpha-1}{a^{1/2}}\nu^2,\quad\alpha\rightarrow1\\
&R_{\rho_{00}}(\nu;\alpha)=2\ln r_{eff}+\ln2\pi+a^{1/2}\left(1-\ln a^{1/2}\right)+\ln\Gamma\left(a^{1/2}+1\right)\nonumber\\
&+\frac{1}{\alpha}\left[a^{1/2}\left(1-\ln a^{1/2}\right)+\ln\Gamma\left(a^{1/2}+1\right)+\frac{1}{2}\ln\frac{\alpha}{2\pi a^{1/2}}\right]\nonumber\\
\label{RenyiAB2_2}
&+\frac{\psi\left(a^{1/2}\!+\!1\right)\!-\!\ln a^{1/2}}{2a^{1/2}}\nu^2+\left[\frac{1}{2a^{1/2}}\!+\!\psi\!\left(a^{1/2}\!\right)-\ln a^{1/2}\right]\frac{\nu^2}{\alpha},\quad\alpha\rightarrow\infty.
\end{align}
\end{subequations}
First, let us point out that at $\alpha=1$ the R\'{e}nyi entropy, Eq.~\eqref{RenyiAB2_1}, turns into its Shannon counterpart \cite{Olendski1}\footnote{Eq.~(38) in Ref.~\cite{Olendski1} contains two typos:  first, the free item "+1" on the upper line of its right-hand side should be dropped and, second, the argument of the function $\psi^{(1)}$ on the third line should be $a^{1/2}$ instead of $a$. Also, the item $\frac{1}{2}$ on the first line of Eq.~(40) there should enter with the negative sign. These typos do not affect any other results presented in that paper.}, as expected. Second, dying coefficient $\alpha$ leads not only to the logarithmic divergence of the entropy but simultaneously suppresses its dependence on the AB field, as Eq.~\eqref{RenyiAB2_0} demonstrates. To exemplify a variation of the speed of change of the entropy with the flux $\partial R_{\rho_{00}}/\partial\nu$ at different R\'{e}nyi coefficients, Fig.~\ref{RenyiABpositionFig2}(a) depicts the quantity
\begin{equation}\label{RenyiABdiff1}
\Delta_\nu R_{\rho_{nm}}=R_{\rho_{nm}}(\nu;\alpha)-R_{\rho_{nm}}(0;\alpha)
\end{equation}
at $a=20$ for $n=m=0$. It is seen that as the parameter $\alpha$ decreases to the extremely small values (eventually reaching zero), the entropy looses its dependence on the flux (eventually becoming completely flat). This has a clear physical explanation; namely, at the vanishing $\alpha$ the integrand in Eq.~\eqref{Renyi1X} degenerates to unity, which is not affected by the variation of the AB field. Increasing R\'{e}nyi coefficient makes the slope steeper and at $\alpha\gtrsim1$ the $R_\rho-\nu$ curve practically does not depend on $\alpha$, as a comparison of the dotted, dash-dotted and dash-dot-dotted lines in the figure reveals. This slope saturation can be also deduced from the analysis of the corresponding terms
$$\frac{1}{2}\left[\frac{1}{a^{1/2}}-\psi^{(1)}\left(a^{1/2}+1\right)\right]$$ and $$\frac{\psi\left(a^{1/2}\!+\!1\right)\!-\!\ln a^{1/2}}{2a^{1/2}}$$ at $\nu^2$ in Eqs.~\eqref{RenyiAB2_1} and \eqref{RenyiAB2_2}, respectively: they almost do not differ from each other, especially at the moderate and large $a$. Let us also point out that the convexity of the R\'{e}nyi entropy and the relation between $R_{\rho,\gamma}$ and Onicescu energy, Eq.~\eqref{Onicescu2}, explains the concavity of the position component of the latter \cite{Olendski1}.

\begin{figure}[H]
\centering
\includegraphics[width=\columnwidth]{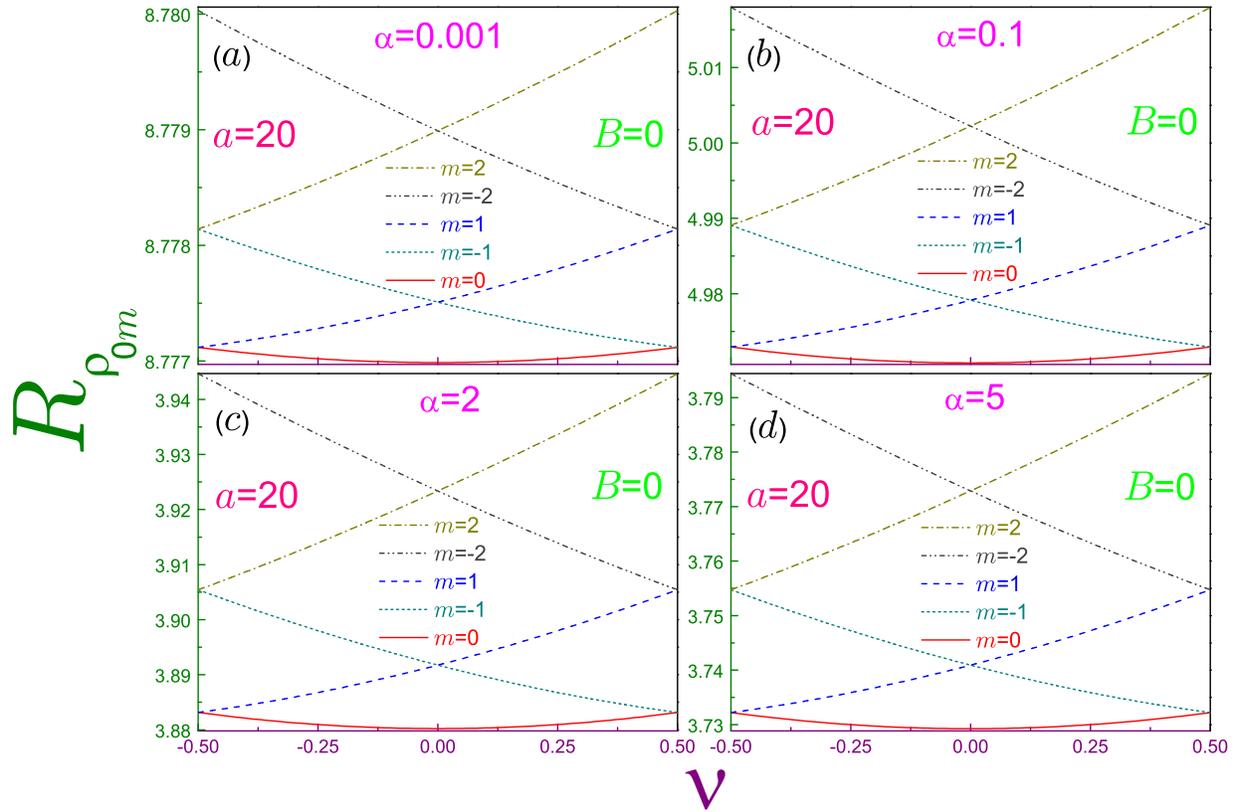}
\caption{\label{RenyiABpositionFig1}
Position R\'{e}nyi entropies $R_{\rho_{0m}}$ as functions of the normalized AB flux $\nu$ at $a=20$, zero magnetic field and several parameters $\alpha$ where panel (a) is for $\alpha=0.001$, window (b) - for $\alpha=0.1$, subplot (c) is for $\alpha=2$ ad panel (d) shows the entropies at $\alpha=5$. In each of the windows, solid curve denotes the orbital with $m=0$, dotted line is for the level with $m=-1$, dashed line - $m=1$, dash-dotted line describes the entropy of the state with $m=-2$, and dash-dot-dotted curve - with $m=2$. Radius $r_0$ is assumed to be equal to unity. Note different scales and ranges of the vertical axis in each of the panels.}
\end{figure}

Fig.~\ref{RenyiABpositionFig1} shows $R_{\rho_{0m}}-\nu$ characteristics for three smallest $|m|$ and several R\'{e}nyi parameters. Corresponding analysis of the Shannon dependencies revealed a strong similarity between $R_\rho(\nu;1)$ and the energy spectrum \cite{Olendski1}. This resemblance survives qualitatively at the arbitrary coefficient $\alpha\neq1$; in particular, relations
\begin{subequations}\label{RenyiAB3}
\begin{align}\label{RenyiAB3_1}
R_{\rho_{nm}}\!\left(-\frac{1}{2};\alpha\right)&=R_{\rho_{n,-m+1}}\!\left(-\frac{1}{2};\alpha\right)\\
\label{RenyiAB3_2}
R_{\rho_{nm}}\!\left(\frac{1}{2};\alpha\right)&=R_{\rho_{n,-m-1}}\!\left(\frac{1}{2};\alpha\right),
\end{align}
\end{subequations}
which are elementary derived from Eq.~\eqref{Renyi2X}, are an exact replica of the corresponding degeneracy of the energy spectrum in the zero uniform magnetic field \cite{Olendski1}. This is a consequence of the invariance of the radial part of the position waveform, Eq.~\eqref{RadialWaveforms1R}, energy, Eq.~\eqref{Energy0}, and persistent current, Eq.~\eqref{Curr1}, under the transformation
\begin{equation}\label{Transformation1}
m\rightarrow m-1,\quad\nu\rightarrow\nu+1.
\end{equation}
Also, at any $\alpha$ the slope retains the same sign as the azimuthal index $m$. Quantitatively, the magnitude of the steepness $|\partial R_{\rho_{nm}}/\partial\nu|$ for any orbital, similar to the $n=m=0$ state, decreases as the R\'{e}nyi coefficient tends to the progressively smaller values eventually becoming perfectly flat at $\alpha=0$ whereas at $\alpha\gtrsim1$ it is almost not affected by the variation of this parameter. Fig.~\ref{RenyiABpositionFig2}(b) shows both these features for the $n=0$, $m=-1$ level. One can say that the decreasing R\'{e}nyi factor increases the density of the position components with its lowest threshold moving higher and in the opposite regime of the huge $\alpha$ the number of the position R\'{e}nyi entropies per unit interval saturates with its bottom being determined by the antidot strength $a$.

\begin{figure}[H]
\centering
\includegraphics[width=0.9\columnwidth]{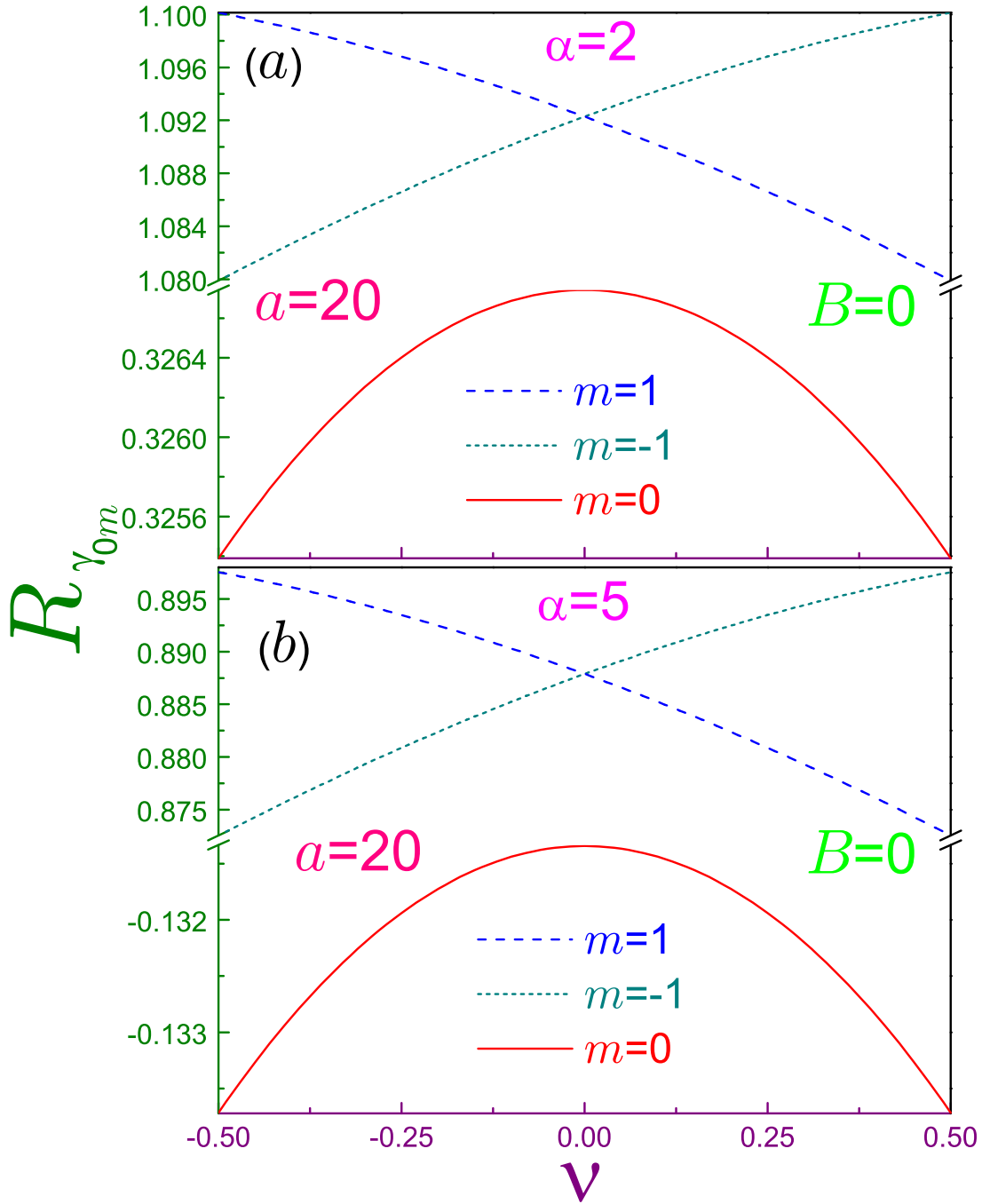}
\caption{\label{RenyiABmomentumFig1}
Momentum R\'{e}nyi entropies $R_{\gamma_{0m}}$ for $m=0$ and $\pm1$ as functions of the normalized AB flux $\nu$ for (a) $\alpha=2$ and (b) $\alpha=5$. All other parameters and conventions are the same as in Fig.~\ref{RenyiABpositionFig1}.
Due to the relatively small change of the entropies as compared to the distance between $R_{\gamma_{00}}$ and $R_{\gamma_{0,\pm1}}$, vertical axes breaks have been inserted in panel (a) from 0.3267 to 1.0799 and in window (b) from -0.13133 to 0.8726. Note also different scales above and below the break in subplot (b).}
\end{figure}

Discussing momentum entropies dependence on the AB field, one has to recall that there is a lower nonzero threshold at which $R_{\gamma_{nm}}$ can be calculated. Eq.~\eqref{Threshold1} reveals that if the momentum component, e.g., for the rotationally symmetric orbitals, $m=0$, takes finite value at the zero flux, it will stay bounded at any arbitrary $\phi_{AB}$. However, the opposite is not always true: a decreasing AB intensity increases for these levels $\alpha_{TH}$ what can lead to the divergence of the corresponding entropy at the fixed R\'{e}nyi coefficient. For $m\neq0$ states the symmetry with respect to the sign of the flux is lost; accordingly, the entropy that was finite at some particular $\alpha$ and zero AB field can become infinite with the variation of the flux. Thus, as mentioned in Sec.~\ref{sec2_2}, the AB intensity can switch the existence of the momentum functionals.

Numerical analysis shows that momentum components $R_{\gamma_{n0}}$, contrary to their position counterparts, are concave functions of the flux. A particular case of this statement for the Shannon entropy, $\alpha=1$, was established before \cite{Olendski1} and is generalized here to all other values of the R\'{e}nyi coefficient. Fig.~\ref{RenyiABmomentumFig1} exemplifies the entropy behavior at the two parameters $\alpha$. Steepness $|\partial R_{\gamma_{nm}}/\partial\nu|$ becomes more precipitous for the larger $\alpha$, as was the case for $R_{\rho_{nm}}$ too. It is observed that for the same orbital the sign of the slope of the momentum R\'{e}nyi functional is just the opposite to its position fellow. The relations similar to Eqs.~\eqref{RenyiAB3} do not exist for $R_{\gamma_{nm}}$ what is a direct consequence of the expression for the corresponding radial waveform, Eq.~\eqref{RadialWaveforms1K}. The gap between the entropies with different $|m|$ gets wider as the R\'{e}nyi factor grows whereas the range of change of each $R_{\gamma_{nm}}$ at $\alpha\gtrsim1$ stays almost unchanged. This is the reason the vertical breaks have been introduced in Fig.~\ref{RenyiABmomentumFig1}.

As a last note of this discussion, let us mention that, similar to the Shannon case \cite{Olendski1}, the background uniform magnetic field, $B\neq0$, does not change the shape of the $R_{\rho,\gamma}-\phi_{AB}$ characteristics but simply shifts them in the vertical direction, as it follows, for instance, from Eq.~\eqref{RenyiAB1}. Accordingly, Eqs.~\eqref{RenyiAB3} representing the invariance under the transformation from Eq.~\eqref{Transformation1} stay intact too. A structure of the energy spectrum in this case is analyzed in Ref.~\cite{Olendski1}.

\section{Conclusions}\label{sec4}
Knowledge of the R\'{e}nyi and Tsallis entropies is important in studying various phenomena in many branches of science. This general fact was confirmed above by showing that, for example, the R\'{e}nyi position components of the QR at any coefficient $\alpha$ qualitatively repeat the behavior of the AB energy spectrum in zero uniform magnetic fields what can be used for predicting the magnitude of the associated persistent currents. Among other findings, let us mention the equation for the lowest boundary of the dimensionless R\'{e}nyi/Tsallis coefficient at which the corresponding momentum components do exist, Eq.~\eqref{Threshold1}, which shows that there is its abrupt jump when the topology of the structure changes from the singly- to doubly-connected one. Note that for the orbitals with their position densities concentrated far away from the origin (what mathematically means that $a\gg1$ and/or $|m|\gg1$), the threshold from Eq.~\eqref{Threshold1} approaches asymptotically that of the QD what physically is explained by the negligible influence of the inner confining potential on their properties. Uncertainty relations for both entropies are independent of the uniform field $B$ and become tight not only for the 2D Gaussians of the lowest QD orbital, Eqs.~\eqref{RadialWaveforms2} with $m=0$, but also they turn into the identity at $\alpha=1/2$ for the QR $n=m=0$ level, which is the only state that reaches this saturation. In this way, earlier conjecture \cite{Olendski3} about the uniqueness of this orbital that should have the lowest energy is amended since the well-known property of the QR energy spectrum is crossings of the levels as the field $B$ increases.

Flexibility of the model described by the potential from Eq.~\eqref{Potential1} leads to miscellaneous limiting geometries \cite{Olendski1,Tan1,Tan2}; in particular, keeping constant the radius $r_{min}=2^{1/2}a^{1/4}r_0$ at which the sole zero minimum of $U(r)$ is achieved and simultaneously unrestrictedly enlarging $\omega_0$, one arrives at the 1D ring of the same radius $r_{min}$ pierced by the total magnetic flux $\phi_{tot}=\pi r_{min}^2B+\phi_{AB}$ \cite{Aharonov2,Merzbacher1,Feinberg1,Peshkin1} when the position waveform, Eq.~\eqref{Representation1_1}, energy spectrum, Eq.~\eqref{Energy0}, and persistent current, Eq.~\eqref{Curr1}, degenerate, respectively, to
\begin{subequations}\label{Rotator1}
\begin{align}\label{Rotator1_Psi}
\Psi_m(\varphi_r)&=\frac{1}{(2\pi)^{1/2}}e^{im\varphi_r}\\
\label{Rotator1_E}
E_m(\theta)&=\frac{\hbar^2}{2m^*r_{min}^2}(m+\theta)^2\\
\label{Rotator1_J}
J_m&=-\frac{e\hbar}{m^*r_{min}^2}(m+\theta),
\end{align}
\end{subequations}
with $\theta=\phi_{tot}/\phi_0$. Observe that due to the frozen radial motion, the principal quantum index $n$ has been dropped from Eqs.~\eqref{Rotator1}. Since $\Psi_m(\varphi_r)$ and $[2m^*E_m(\theta)]^{1/2}$ describe the eigenstates of the angular momentum of this AB rotator, the corresponding R\'{e}nyi uncertainty relation is saturated by them and does not depend on $\alpha$ and $\beta$ \cite{Bialynicki2}. Let us also note that this model apparently can be used as a foundation of the quantum-informational analysis of the relevant more complicated structures, such as, for example, nanohelices \cite{Tinoco1,Kibis1,Vorobyova1,Downing1}.

Armed with the expression for the R\'{e}nyi entropies, one can build up shape R\'{e}nyi complexities \cite{Antolin1}:
\begin{equation}\label{Complexity1}
C_{\rho,\gamma}(\alpha)=e^{R_{\rho,\gamma}(\alpha)}O_{\rho,\gamma},
\end{equation}
where the formulas for the  disequilibria $O_{\rho,\gamma}$ are given in Eqs.~\eqref{Onicescu1}; for example, this was very recently done for a noncommutative anisotropic oscillator in a homogeneous magnetic field \cite{Nath1}. Regarding this dimensionless quantity, let us just to point out that for our geometry neither its position $C_\rho$ nor wave vector $C_\gamma$ component depends on the uniform intensity $B$.

Finally, let us remark that above the R\'{e}nyi and Tsallis functionals were considered in the position and momentum spaces, which are two non-commuting observables. In the last year or so, R\'{e}nyi \cite{Coles2,Rastegin1} and Tsallis \cite{Rastegin1} entropies were proposed in energy and time domains; in particular, corresponding uncertainty relations were derived \cite{Coles2,Rastegin1}. Application of these measures and associated inequalities to the analysis of the QDs and QRs may present an interesting development of quantum information and quantum cryptography protocols.

\vspace{6pt} 

\funding{This research was funded by the Research Funding Department, Vice Chancellor for Research and Graduate Studies, University of Sharjah, SEED Project No. 1702143045-P.}

\conflictsofinterest{The author declares no conflict of interest.} 

\abbreviations{The following abbreviations are used in this manuscript:\\

\noindent 
\begin{tabular}{@{}ll}
AB & Aharonov-Bohm\\
$n$D & $n$-dimensional\\
QD & Quantum Dot\\
QR & Quantum Ring
\end{tabular}}

\reftitle{References}

\end{document}